**An Income-Based Approach to Modeling Commuting Distance in the Toronto Area**

Shawn Berry, William Howard Taft University

## 1.0 Introduction

The Greater Toronto Area is, without question, a major housing market that has seen continuous overheating and a resulting exodus of homebuyers in search of affordable housing beyond the traditional suburbs of Toronto. Canada Mortgage and Housing Corporation (CMHC) refers to this phenomenon as "drive until you qualify" (CMHC, 2018), and as a type of migration of homebuyers who are "being priced out of the Toronto market, move further and further into the suburbs until they find housing they can afford (and a mortgage they can qualify for)" (CMHC, 2018). This phenomenon has also created a large strain on household budgets which may not be sustainable. In fact, research from the Bank of Canada (2018) has reported that households are holding mortgages that have loan values which are 4.5 times the annual income of the households, which has led to what is referred to as a mortgage stress test.  Mortgage stress measures the ability of a household to carry shelter costs and debt relative to household income, and sets an allowable maximum debt ratio for mortgage applicants. However, homebuyers are searching further afield for less expensive housing in order to meet this mortgage qualification criteria. As shelter costs skyrocket in large urban centres such as Toronto, this exodus to areas beyond the suburbs has resulted in longer distances to work that rely on cars, with over 72% of commuters using cars to travel from distances of 20 km or greater. (Deloitte, 2016).

This raises the question of how far a household would be willing to commute but not exceed the mortgage stress guidelines. It is important to note that the change in residential location being considered in this paper is not as a result of a workplace relocation but as flight from high shelter costs



to areas of lower costs. Thus, we are assuming that the household keeps their workplace location static (i.e., Toronto) and is willing to commute. We will look at creating a model that is based on the cost of driving, the shelter cost as represented by a regional curve for Toronto, and a percentage of household income that is allocated to shelter and transportation to estimate a distance from Toronto which does not exceed a budgetary constraint. Alternatively, given a distance from Toronto, we can compute a percentage allocation of household income and compare it to a given guideline for spending to determine if a household can sustain this level of spending, or may be tempted to search further afield. The model will reveal boundaries of feasible (i.e., costs are less than income allocated) and infeasible (i.e., costs are greater than income allocated) housing locations which ultimately exclude consideration of housing within or outside of these limits. Furthermore, with shifts in shelter and transportation costs over time, the model will illustrate how these boundaries of feasible and infeasible housing locations will change and affect commuting distance as a result.

The significance of this model is the novel approach to evaluating household location decisions based on shelter and transportation costs, subject to a budget, in an elegant way using only distance as an independent variable. The theoretical underpinnings of this model is inspired by the work on the classical rent gradient by Alonso (1964), Muth (1969), and Mills (1967). However, the practical significance of the model presented here is the minimal number of assumptions used as a rule for determining commuting distance as a function of budgeted income. Variables in the model can be immediately operationalized with available data to estimate the effects of changes in a housing market, income allocation, income level, or transportation mode on residential location and commuting distance. Furthermore, the model can be modified to evaluate different transportation mode choices or combinations using the same equation form. The model can be rearranged to evaluate affordability (i.e., percent of household income), or level of required household income, given other details about shelter and transportation mode.



In the section that immediately follows, this paper will discuss the literature surrounding the relationship between commuting distance and housing affordability in the context of mortgage stress, and the allocation of income to transportation and housing costs. Next, we will introduce the model and its specification. The findings will be presented wherein the model is estimated and used to illustrate the effects of changes in shelter costs on commuting distance in the Toronto region from 2011 to 2016. The implications for managerial practice will be discussed after the presentation of the results and findings. Finally, we will discuss the limitations, and conclude the paper.

## 2.0 Literature Review

The rent-bid function (Alonso, 1964; Muth, 1969; Mills, 1967), which illustrates a relationship between rent cost and distance, has been used as the basis to develop models that examine commuting behavior in the context of the relationship between residential location and the workplace. The nature of the rent-bid function is said to be negatively-sloped and convex as distance increases from a central city (Anderson, 2008). Yinger (2005; 2020) has also used the rent-bid function to explain access to employment and the relationship between housing and commuting costs. While Simpson (1987) indicates that some models poorly explain the workplace-residence relationship, Simpson does confirm that models also generally do not deal well with fixed residential locations. However, recent work from Yinger (2020) affirms the use of the rent-bid function as a valid approach for commuting analysis. In the context of this paper, the rent-bid approach will be used to frame the drive until you qualify problem in terms of limiting commuting distance according to household income.

The measurement of commuting distance is extensively discussed in Hu & Wang (2016; 2020), specifically time versus mileage, and the nuances of how distance should be measured to reflect the



actual commuting distance in a non-biased way. While Hu & Wang (2016) state that mileage is a more preferred method of measuring commuting distance versus time, they state that "more accessible and accurate measures of commute length remain very much needed" (p.6). Since Google Maps was used as a method for determining driving distance between Toronto and other communities in this paper, it is worthy to note the work of Hu et. al (2020), and their use of Google Maps to analyze driving times and distance between origins and destinations zip codes. With respect to the use of Google Maps, Hu et. al (2020) state that "researchers usually use this approach to derive drive times for a limited number of OD pairs" (p.3), but caution that Google Maps "cannot measure travel distances or times from a place to itself" (p.6). For this reason, the distance used from Toronto to itself was assumed to be 10 kilometres for convenience in this paper.

The relationship between commuting and household income has been examined by various authors, all generally characterized in the same way. Carra et al. (2016) state that longer commuting distances will be observed with increasing income, and generated a model that predicts the average commuting distance using income and transport costs. Park and Quercia (2015) state that shelter costs and commuting costs are traded off by households, observing that shelter costs declined with distance as predicted, and that longer distance commuting was observed at higher levels of income. Kellett et al. (2016) observes "where there is a lack of affordable housing, there is evidence of trading of transport cost for housing cost, for example, as households seek more distant but lower cost locations for their residential location" (p.2). Orazem and Otto (2001) similarly observe that "individuals make residential and job location choices by trading off wages, housing prices, and commuting costs" (p.1036). Ahrens and Lyons (2021) also allude to the trade-off between high housing costs and commuting times (as a proxy for distance), stating "some commuters might be forced to take on longer commutes due to rising rents in central locations" (p.264), noting that an increase in housing costs in an employment centre resulted in an increase in commuting time. The observation by Ahrens and Lyons



(2021) is important, given that the dynamic of increasing shelter costs is clearly a driver for increased commuting distance.

Schuetz (2019) characterized the relationship between shelter costs, commuting, and the proportion of income as mortgage stress, created particularly in expensive urban markets, and resulting in longer commutes. Schuetz (2019) also points out the disparity in homeownership rates between higher and lower income quartiles and that homeowners and renters making $30,000 a year or less spend more than 30% of their income on housing. The popular 30% rule of thumb is also mentioned in Luffman (2006) in a discussion on affordable housing in Canada. However, the percentage of income being spent on transport and housing must be considered. This percentage is referred to in the literature as a H+T (housing plus transport) burden (Coulombel, 2018; Guerra & Kirschen, 2016). Specifically, Coulombel (2018) states that "several works report housing plus transport (H+T) burdens increasing with distance to the city center" (p.90). Coulombel (2018) notes that "high H+T burdens are likely to increase the risk of insolvency" (p.90), underlining the importance of transportation costs in this burden measure. Specifically, Guerra & Kirschen (2016) state that "a neighborhood (is) affordable if a given household would spend 45% or less of its income on housing and transportation costs" (p.7). A 2018 report by the Canada Mortgage and Housing Corporation specifically looks at the dangers of the trade-off between moving outward beyond the suburbs for cheaper housing and car commuting, which they refer to as drive until you qualify. Mitra & Saphores (2019) comment that "few papers have considered how housing costs influence long-distance commuting" (p.1). This underlines the importance of the current paper.

Despite the attention given to the rise in drive until you qualify home buying and increasing housing prices in the Toronto area, there is a sparse recent literature in a Canadian context to understand the relationship between household income, residential location, shelter affordability, household expenses



and transportation costs. Kellett et. al (2016) used Australian census data for 2006 and 2011 to examine changes in commuting as a result of drive until you qualify in metropolitan markets based on the proportion of income spent on shelter and driving. Kellett et. al (2016) state "Hanson, Schnier et al. (2012) define the "drive-'till-you qualify" (DTQ) condition as follows: 'the credit constrained household locates as far out as it must to afford the quantity of housing closest to what would be its unconstrained demand, balanced against the incremental commuting costs'" (p.2). This credit constraint may result in homeowners taking on large amounts of debt just to purchase a home. In particular, where homeowners have taken on a lot of debt (i.e., high leverage), Rouwendal (2014) found that these homeowners tend to have longer commutes, which validates the observation by Kellett et al. (2016). Keil (2018) mentions the phenomena in passing during his discussion on suburbanization by saying that "once there was the formula "drive until you qualify" that set the boundary for homebuyers in a given region." While Keil (2018) did not discuss this further, his reference to a "boundary for homebuyers" is an outcome of the model being presented in this article, and will help define a set of boundaries for zones of affordability within an urban region. Luckey (2018) attempts to examine the relationships between household income and shelter costs and other household expenses (e.g., child care) to explain variations in residential location, but does not attempt to define limits of viable residential location. While Newbold and Scott (2013) examine the commuting distance of migrants in the Golden Horseshoe within the context of urban sustainability and Ontario's *Places To Grow* and Greenbelt urban policies, they acknowledge the phenomena by reporting a trade-off of longer commuting distances in exchange for lower shelter costs.

Given that drive until you qualify is a symptom of households that must relocate further away from a workplace because of credit or income constraints (Hanson, Schnier et al., 2012), this differs from the concept of residential self-selection that considers factors that influence the preference of residential locations since "choices are mostly constrained by what they can afford in combination with the



availability of housing in their preferred neighborhoods" (Lin et al., 2017, p.113). Citing Frank et al. (2007) and Nass (2005), Lin et al. (2017) remarked that "safety and housing price seemed to be more important than travel-related preferences when deciding where to live" (pp.112-113). This is not surprising, given the trade-off of lower housing costs versus higher commuting costs that is commonly cited in the literature.

Since drive until you qualify suggests a household constraint that forms a boundary for homebuyers that trade lower housing prices for increased commuting distances, it is logical to consider the literature surrounding commuter sheds.  Axisa, Newbold and Scott (2012) created a model of commuting for migrants in the Toronto area but "individuals with commute distances in excess of 100 km were excluded to avoid including weekly (long‑distance) commuters" (p.36). While Axisa, Newbold and Scott (2012) do suggest that "households are forced to strategise commuting and residential location choices around family, work and residence location" (p.41), their work deliberately excludes census subdivisions outside of their desired study area within the Toronto commuter shed (Axisa, Newbold and Scott, 2012, p.346). This appears to be a shortcoming to understanding the extent of long-distance commuting within the Toronto commuter shed, which they state is "an area that closely corresponds to the Greater Golden Horseshoe (GGH) region of Ontario" (Axisa, Newbold and Scott, 2012, p.345), especially given the growth pressures that they acknowledge immediately after this sentence, and the vast scale of the commuting region itself. This paper will demonstrate that car commuters are indeed motivated to drive greater than 100 kilometres due to an intense search for lower shelter costs, implying that the Toronto commuter shed extends much further in distance due to the drive until you qualify phenomenon that is driven by raising shelter costs.

While Luckey (2018) looks at budgetary considerations in a study of transportation decisions as they relate to housing affordability, the author observes "there has been surprisingly little critical



examination in either the academic or public spheres around how to define and measure the concept of 'affordable housing'" (p.37). In the context of drive until you qualify, there does not appear to be literature which specifically looks at the percentage of household income allocated to shelter and transportation as a function of distance, subject to budgetary constraints (i.e., affordability) which could help someone determine where a given commuting distance might actually be detrimental to a household budget. The model presented in this paper will attempt to speak to this household budgetary constraint, and frame it in terms of commuting distance that avoids mortgage disqualification or burdening the household with high shelter and transport costs.

While the model presented here seeks to generalize the macro effects of increases in shelter prices in a metropolitan or regional housing market on car commuting distance (i.e., long distance commuting), subject to income levels and income constraints, the model could help form a framework to evaluate intra-zonal commuting (i.e., short distance commuting) using changes in shelter costs. Hu and Wang (2016), in their analysis of intraurban commuting in Baton Rouge from 1990 to 2010, provide insights into the roles of time, distance, land use layout, and wages on wasteful commuting. Hu and Wang (2016) discussed the shortcomings of the estimation of commuting distance, revealing that it is subject to bias and overestimation when done for intra-zonal trips. While Hu and Wang (2016) did not examine the changes in commuting distance as a function of temporal changes in shelter costs, the model presented here could be potentially used to help examine changes in intrazonal commuting with changes in costs (i.e., shelter and transport) and allocation of income using this approach.

With regard to the mortgage stress test, the costs considered are not just shelter costs which include taxes among other things, but also debt servicing costs, which refers to the amount of household debt being carried for debt of every kind (CMHC, 2018). The CMHC refers to the Gross Debt Servicing (GDS) ratio, which is the ratio of the total of debt principal plus interest plus taxes plus heat, and



divided by gross annual income; GDS should not exceed 35% (CMHC, 2018). The CMHC also refers to another critical metric in the mortgage stress test called Total Debt Servicing (TDS) ratio (CMHC, 2018). TDS is equal to the total of debt principal plus interest plus taxes plus heat plus other debt obligations, and divided by gross annual income. To qualify for a mortgage and pass the mortgage stress test, the overall shelter cost plus debt servicing should not exceed 42% of the household gross annual income (CMHC, 2018).

**3.0 Research Question & Methods**

*3.1 Model Specification.* Borrowing from a popular concept in microeconomic theory that deals with the costs of a firm and the short-run equilibrium in perfect competition (Arrow & Debreu, 1954) and optimality (Lipsey and Kelvin, 1956), a similar representation can be constructed for commuting and shelter costs, subject to a simple allocation of income to housing and driving costs. In fact, Brueckner (2000) follows a very similar way of examining the economics of urban sprawl by graphically representing average and marginal costs over distance in the same manner economists map costs of a firm over increasing amounts of production, and finding optimal solutions based on utility. Unlike Brueckner (2000), the model being presented here is concerned with how far out a commuter ought to stop in search of housing before exceeding a certain income allocation guideline, given an income, and is not based on utility maximization. It is being presented using a similar analysis framework that one would graphically use in microeconomics. We can establish the lower and upper bounds where a commuter can feasibly commute from based on income constraints (i.e., mortgage stress guidelines, rules of thumb), and given after-tax household income and the total cost of shelter and transport to work from Toronto. Using the shelter cost, total cost and transport cost curves, this simple condition is satisfied as follows:



$$TC(d) \leq p*I$$

where $p$ is the proportion of after-tax household income spent on transport and housing (i.e., income constraint), $I$ is the after-tax household income and, $TC(d)$ is the total cost of living in the community and traveling to work in Toronto (instead of living in Toronto), a function of one way distance from the community to Toronto in kilometers. This income constraint specifies that the left side of the equation (i.e., total cost at a given distance) must be less than or equal to the right side of the equation (i.e., budgeted income) for the location to be feasible. This presupposes that $p$ is known and fixed at some level (e.g., mortgage stress guideline of 42% of income to qualify for a mortgage). The model assumes that homebuyers work in Toronto and commute by car driving, giving consideration to possible housing location choices that include Toronto and beyond its borders. Values of after-tax household income, shelter costs and transport costs use monthly data for ease of calculation. The beauty of this model specification is that it can be estimated quickly and easily using Census data, and relies on a simple rule that is based on the budget restrictions set by the CMHC. However, one could use any maximum budget constraint of their choosing in the model (e.g., 45% of household income).

While driving time (i.e., minutes) was given consideration as a possible measure for distance in the model, the mileage distance (i.e., kilometres) was used for a few reasons. First, the concept of mileage is better understood since it is fixed between two points whereas the driving time can change over time for different reasons (i.e., traffic conditions, route speed limits can change). Second, Hu & Wang (2016) recommend the use of mileage instead of time as a preferred method of distance measurement. This will help to model costs in the context of location as measured by the number of kilometres from Toronto, and reveal the shape of the regional cost curves over distance. Finally, since driving cost data using secondary sources was expressed in terms of distance (e.g., litres per 100 kilometres, cost per



kilometre), the use of driving time as a distance measure was excluded for this reason. However, driving time data for each community are presented alongside the driving distance data to provide additional context.

Figure 1 illustrates the housing market in terms of zones where housing is not affordable, transport is not affordable, and a zone of feasible locations that are within the income constraint limits. The graphical representation of costs will appear familiar to those who have studied economics, with the x-axis showing distance from Toronto and the y-axis showing dollars (cost and income). The shelter costs are mapped as curves and we want to examine the characteristics of the market for a commuter relative to a horizontal budget line. At **d1**, this is the distance at which shelter costs (*R(d)*) equal the percentage of after-tax household income allocated (*p\*I*), and it forms the boundary for the zone of high housing cost. At distances less than or equal to **d1**, the allocation of income is insufficient to enable living in this zone when *R(d)>p\*I*. At **d2**, this is the distance at which transport costs by car (*T(d)*) equal the percentage of household income allocated, forming the outer boundary of feasible housing locations. At distances greater than **d2**, transport cost will be greater than the income constraint (*T(d))>p\*I*)**. Between points **d1** and **d2** is the zone of feasible housing locations which satisfy the constraint on allocated income, which we could call the *Goldilocks* zone - not too expensive, but just right. The intersection of the shelter cost and transport cost curves is noted as an indifference point, where shelter costs equal transport costs. We will also note that the rent curve begins an upslope from a minimum at some distance past Toronto, where observed shelter costs begin to increase. This suggests the influence of an adjacent major metropolitan housing market (e.g., Ottawa) on communities beyond the Toronto metropolitan housing market. Figure 1 could be generalized to major urban housing markets. We understand that the first limit of feasible locations, **d1**, represents our starting point where households have accepted that living in Toronto is not affordable and must look further afield because it is not sustainable to support these expenses, even with modest incomes.



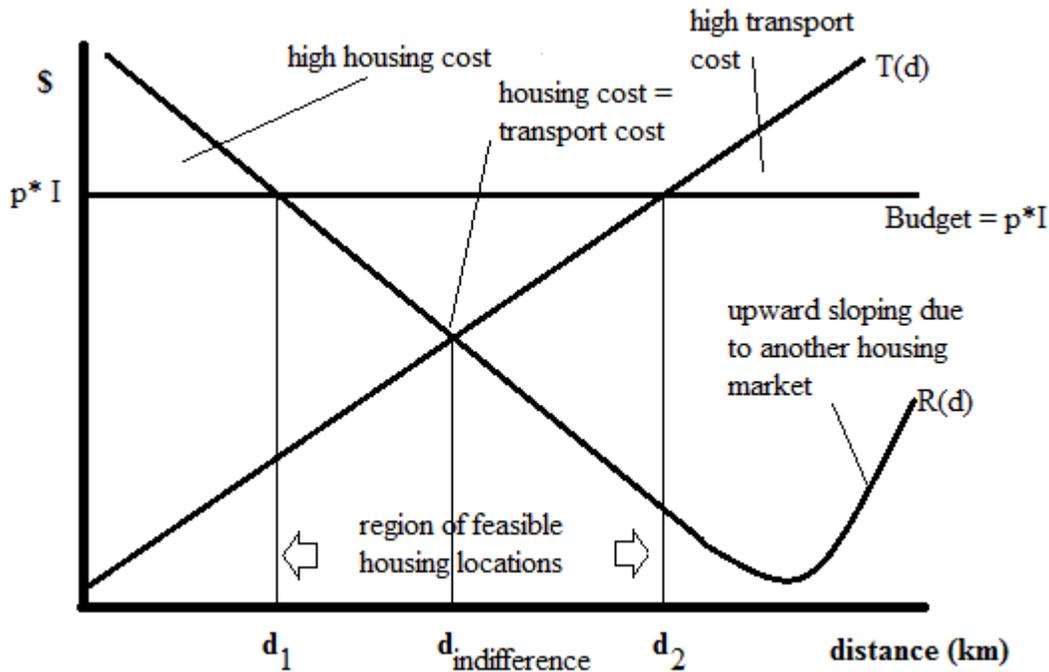

**FIGURE 1: Rent curve, transport costs and income allocation as a function of distance**

**Source: generalized from model calculations (not to scale)**
(require black and white, 1.5 column)

*3.2 Research Hypothesis*    We make the assumption that the household maintains a desired

percentage allocation of income $p$ to live at some distance $d$. This said, there will be some distance

beyond which the household may not be willing to live past except under one of two circumstances.

The first circumstance is if the household decides to allocate more of their budgeted income (e.g., from

40% to 50%) to secure shelter that is less expensive but located further away. The other circumstance is

if the household simply begins to earn more income and can now choose to live further away while still

maintaining a proportion of household income devoted to shelter and transport that they are

comfortable with (which may be more or less than before).

The total cost of shelter and transport, **TC(d),** at a given distance, d, is specified as follows



$$TC(d) = R(d) + T_1(d) + T_2(d) + PC$$

where $R(d)$ is the shelter costs at a distance (in kilometres) from Toronto, $T_1(d)$ is the transport cost by car, $T_2(d)$ is the transport cost by a second mode (e.g., bus, transit), and $PC$ is parking cost. This model is applicable to any given region, and not unique to Toronto, making it widely applicable and easy to construct.

As seen in Figure 1's generally downward sloping shelter cost curve, we assume a traditional rent-curve which we will refer to as the shelter cost curve where we expect less expensive shelter costs with increased distance from the community where work is located (i.e., negatively correlated with distance.) $T_1(d)$ is assumed to be the cost of driving by car to work in Toronto, which clearly increases with distance but is dependent upon a number of cost assumptions (i.e., average litres per 100 kilometres, average price of gas, etc.). If no other mode is involved (e.g., train, subway) for the cost of the second mode ($T_2(d)$), then this value will be zero. Otherwise, $T_2(d)$ is the total cost when a second mode of transport is required in the workplace city, like local transit, and does not have to be formulated as a function of distance (e.g., can be a monthly transit pass amount). An important implication of $T_1(d)$ and $T_2(d)$ in this model is that when they are equal (i.e., monthly driving cost equals the monthly cost of some other mode), the commuter should be indifferent to the use of either car or transit. If one cost is less than the other, the commuter should favor the less expensive mode. If a commuter drives to a transit terminal or train station, the value of $d$ in $T_1(d)$ simply becomes the one-way driving distance to that point where the car is parked and transit is used. $PC$ may also be zero if the commuter has no parking costs, or otherwise be the total monthly cost for parking at the workplace city, or at some intermediate point where the commuter changes from car driving to some other mode to get to and from work (e.g., drives from Peterborough to Oshawa GO Train station, drives



from Colborne to VIA Train station in Cobourg.)  In the case of this article, $T_2(d)$ and $PC$ were set to

zero as we wish to look at drive until you qualify commuters who use car-only transport, and determine

the boundaries of feasible housing locations for the region using our income constraints.

*3.3 Ranges of the Values of percentage of household income (p)*.    As Table 1 shows, the total

percentage of income spent on shelter and transport (***p***) appears to vary by location (i.e., city,

metropolitan region, and province) and demographic group (i.e., married couples, single parents, etc.).

**Table 1: Sample values of total percentage of spending on housing and transport in Canada**
**Sources:  Statistics Canada, Survey of Household Spending, Table 1, Average spending on goods**
**and services and shares of spending of major categories by Province (2012)**
**Statistics Canada (2009) Average Household Expenditures, By Selected Metropolitan Area,**
**Cansim Table 203-0001, Catalogue # 62F0026MIE**

| Location | % Shelter | % Transport | *p (sum of % Housing and % Transport)* |
|---|---|---|---|
| Canada – all | 28.10% | 19.90% | 48.00% |
| British Columbia | 30.80% | 17.50% | 48.30% |
| Ontario | 29.50% | 19.80% | 49.30% |
| New Brunswick | 22.30% | 24.50% | 46.80% |
| Toronto CMA | 20.50% | 12.70% | 33.20% |
| Montreal CMA | 19.50% | 13.00% | 32.50% |

The value of ***p*** can also be shown as a variable that represents the percentage spent on a combination of

housing and transport for a given socioeconomic group (eg: low income young adults, rural job

seekers, etc.)  This percentage changes with income quintiles. In the lowest income quintile,

households spent 34.8% and 15.1% on shelter and transportation (total 49.9%), respectively;

households in the highest income quintile spent 27.4% and 21.1% on shelter and transportation (total



48.5%), respectively (Statistics Canada, Survey of Household Spending, 2017). Although both quintiles arguably spent about the same total proportion of income, the higher income quintile spent a lower percentage on shelter but a higher percentage on transportation. This illustrates the frequently cited trade-off condition in the literature wherein higher transportation costs are substituted for lower shelter costs. The average percentage spent by Canadians on housing was 29.2% and 15.9% on transportation (Household Spending Survey, 2017). Residents in Ontario spent 31.1% on shelter (the highest percentage in Canada) and 19.6% on transportation (Household Spending Survey, 2017), for a total of 50.7% allocated to these two spending groups. Recall earlier that the CMHC restricts the percentage of income spent on shelter and debt cost, without consideration of commuting cost, to 42% of annual gross income. While household spending on shelter in Ontario is the highest level in Canada, the fact that Ontario residents are allocating more than 50% of income on shelter and transportation should be troubling for anyone using the 42% threshold of Total Debt Servicing. Given this constraint, this raises the question of what should be a prudent level of spending when it comes to commuting costs. The behavior in this case suggests that people may be exceeding a recommended budget allocation to live at a given distance, if an increase in household income is not feasible (i.e., household income being held constant.) Therefore, the maximum allocation of household income to shelter and transportation costs is assumed to be no more than 45% in order to limit the possibility of unaffordability, as recommended in the literature (Guerra and Kirschen, 2016). From Table 1, it is clear that many jurisdictions in Canada exceed the 45% guideline, and would be considered unaffordable by this standard.

*3.4 Values of T(d).*     Since the model is based on commuting by automobile, *T(d)* is the total monthly cost to drive based on a daily round trip between home and work, which has a one-way distance of *d* kilometres from Toronto. Unfortunately, Statistics Canada does not publish driving costs as part of its *Survey of Household Spending*, nor is there published Census data on car commuting costs for a given census metropolitan area or census agglomeration. Worthy of note, despite the research into



the journey to work by Statistics Canada, the measurement of transportation costs could have brought greater context to their collected commuting data. A general lack of transport cost data was also observed by Coulombel (2018), stating "moreover, all works are limited by the unavailability of households' transportation budgets in home loan datasets, so that some accessibility measure (walk score, gravity index, number of vehicles…) is used as a proxy instead" (p.90). The lack of commuting cost data at the census metropolitan area and census agglomeration levels is clearly a challenge for anyone seeking to examine and compare car commuting costs for any given Census year. Therefore, it becomes necessary to estimate **T(d)** based on available driving cost information. Haas et al. (2013) used driving cost data from the American Automobile Association (AAA) in their computations of household transportation costs.  In this paper, to overcome the lack of commuting cost data at the census level, car driving cost data from the Canadian Automobile Association (CAA) (combined with other secondary data sources) for 2011 and 2016 were used to arrive at estimated annual driving costs for commuting from a given distance **d** kilometers from Toronto. This driving cost can be computed as

$$T(d) = gas(d) + insurance + licence + depreciation(d) + finance + maintenance + tires(d)$$

where the values for each cost component for 2011 and 2016 are given in the Table 2 below.

**Table 2: Transport model assumptions for 2011 and 2016 driving costs**

**Sources: CAA Driving Costs 2011 & 2013**

| Driving Cost Component | Definition and Source | Values |
|---|---|---|
| *d* | one-way driving kilometres to Toronto per Google Maps, annualized as follows: (d*2)*5*52 | -- |
| *gas(d)* | gas cost as a function of distance, calculated | 2011 average fuel price: $1.29/l |
| | as follows: (d/100)*litres per 100 kilometres * average fuel price for a given year. Sources: iea.org, canada.ca | 2016 average fuel price: $1.02/l 2011 average l/100 km: 8.0 2016 average l/100 km: 7.3 |
| *insurance* | annual insurance cost for mid-sized sedan, per CAA Driving Costs (Honda Civic LX) | 2011 insurance cost: $1,936 2016 insurance cost: $2,630 |
| *licence* | annual licence cost, per CAA Driving Costs | 2011 licence cost: $115 2016 licence cost: $146.16 |
| *depreciation* | annual depreciation cost as a function of distance, calculated as follows: $28/1,000 km if over 18,000 km per year, otherwise $3,515 per year, per CAA Driving Costs | 2011 depreciation cost and 2016 depreciation cost: $28/1,000 km if over 18,000 km per year, otherwise $3,515 per year |
| *finance* | annual finance cost, per CAA Driving Costs | 2011 finance cost: $699 2016 finance cost: $836.64 |
| *maintenance* | annual maintenance cost per kilometre, per CAA Driving Costs. Calculated by multiplying annual kilometres as defined above by the applicable rate | 2011 maintenance cost: $0.0243/km 2016 maintenance cost: $0.0327/km |

*3.5 Values of R(d).*     The cost for shelter at a distance $d$ kilometres from Toronto is represented by $R(d)$. This can be a function (i.e., shelter cost curve) or an individual value if evaluating based on an average cost of housing. $R(d)$ is specified in the form of a downward-sloping rent curve function:

$$R(d) = R_0 - \beta d + e$$

where $R_0$ is the intercept value for the  curve, $\beta$ is the slope coefficient on the distance variable, and $e$ is a random error variable. We expect shelter costs to be negatively correlated with increasing distance ($d$) from Toronto, where shelter costs are traditionally the highest, and resembles the classical rent-bid function as illustrated in Anderson (2008) and Yinger (2005; 2020).

While there are several sources of data for shelter costs, data from the 2011 and 2016 Census of



Canada were used since this amount includes household expenses related to homeownership. As a comparison of alternate data sources, for example, Padmapper.com can be used as a source for rental market information, and cost curves can be easily constructed using this specification for any urban area. The advantage of these data sources is the ability to construct a large sample for a regional housing market and model it based on attributes, such as number of bedrooms or if the housing is a high-density residential building. However, this can be complicated by the fact that the data contains listings that are aggregated from other sites, such as Craigslist.com, where data quality may be questionable, and introduces the problem of bias as a source of error.

To evaluate changes in shelter costs over time, the shelter costs for the 2011 and 2016 Census years were modelled as a function of distance from Toronto using the previously specified rent-bid function. Data for census metropolitan areas and census agglomerations were used for Ontario communities as a proxy for the overall Toronto region housing market. Communities that were located extremely far from Toronto were excluded (e.g., Arnprior, Brockville, Carleton Place, etc.) since it was unlikely that anyone would be commuting to Toronto on a daily basis by car from these places. As a guidepost for this exclusion, the number of lone driver commuters to Toronto was examined and a distance of 156 kilometres was used to limit the dataset (see Table 3) since this was the distance where the number of lone drivers was negligible. When the 2011 Shelter Costs for 32 of Ontario's 36 census metropolitan areas and census agglomerations were plotted against distance from Toronto, the shelter cost curve reaches a minimum at a distance of 250 kilometres from Toronto (near Kingston) and then begins an upward slope (see Figure 2). This suggests the influence of another adjacent major housing market on average shelter costs (e.g., Ottawa), and is perhaps not unexpected, given that the rent-bid function is said to be convex (Anderson, 2008).



**FIGURE 2: 2011 Shelter costs versus kilometres from Toronto, n=32 CMAs and CAs**

**Source: Census of Canada 2011**
(color required, 1.5 column)

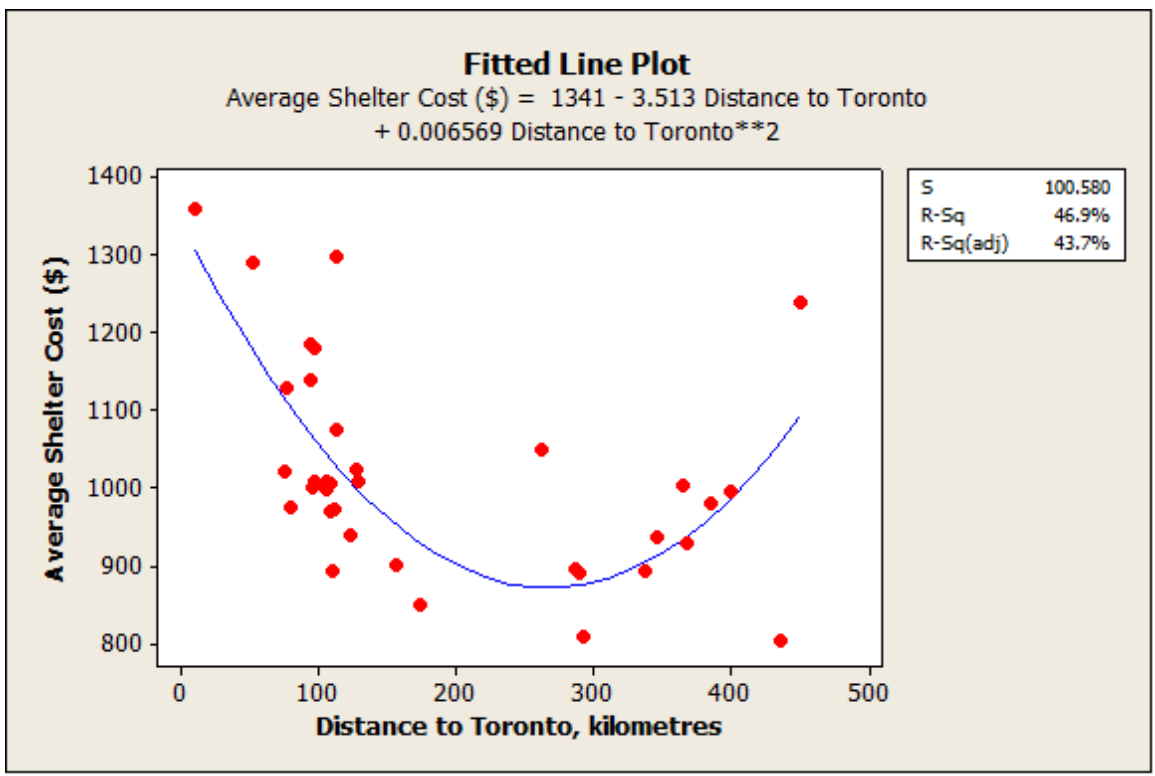

The drive time polygon map in Figure 3 provides a geographical context to the distance and time data in Table 3 for the Toronto region and beyond. Given the observed number of lone car drivers that commute to Toronto (see Table 3) at distances beyond 156 kilometres from Toronto, this suggests that



these commuters may be willing to spend more on transportation costs to live more than two hours away from Toronto. The model presented here suggests an explanation for the existence of these observations of car commuters is due either to elevated levels of household income, or allocating a higher percentage of household income to shelter cost and transportation than expected.

**FIGURE 3: Drive time polygons - 30, 60 and 120 minutes from Toronto by car**
**Source: created at oalley.net**
(black and white or grayscale required, 1.5 column)

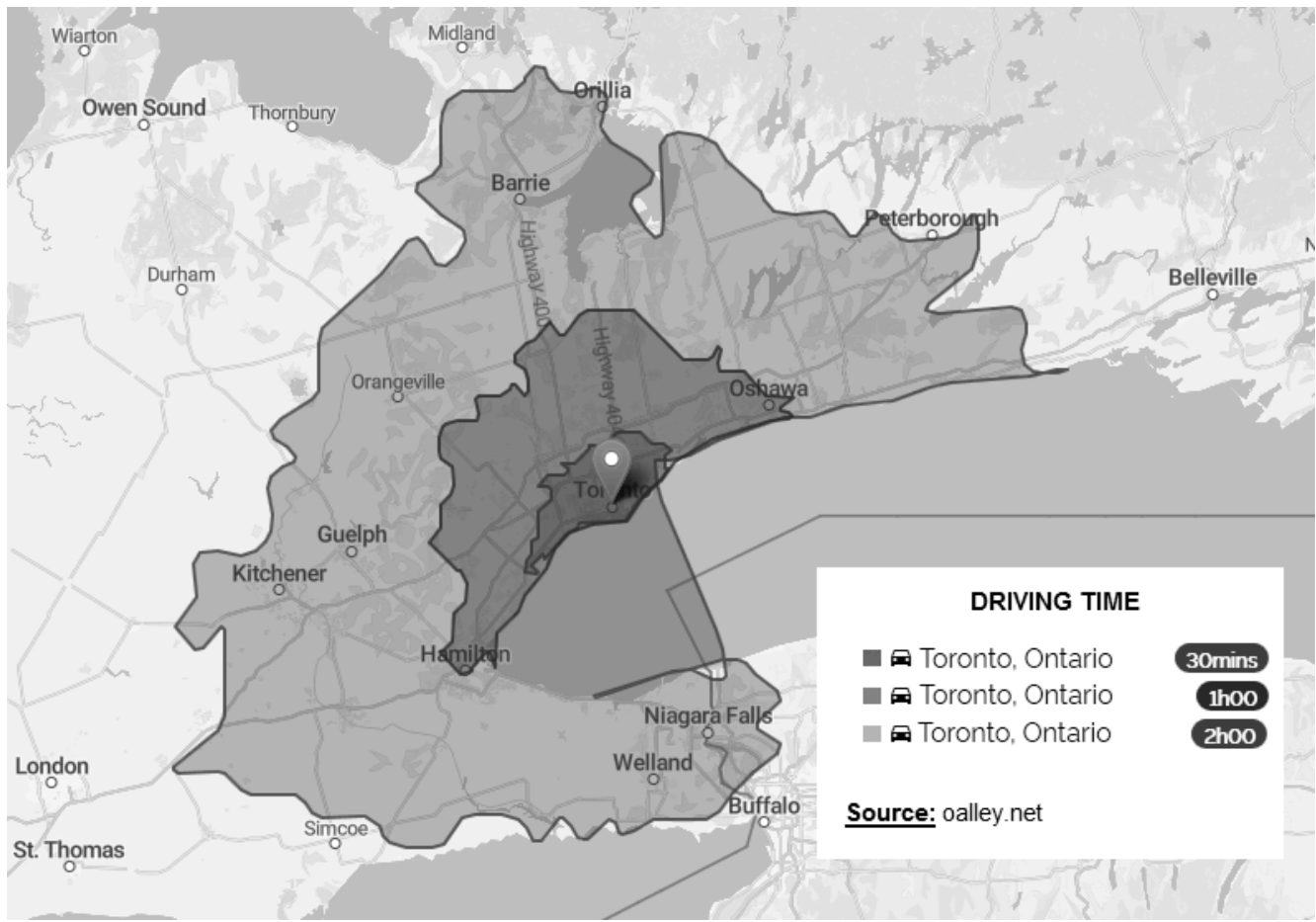



Table 3 illustrates the sample of 23 Ontario Census Agglomerations and Census Metropolitan areas.

**Table 3: Sample of Household Income, Shelter Cost, Driving Costs and Lone Drivers - Ontario Census Agglomerations (CA) and Census Metropolitan Areas (CMA), 2011 and 2016**

**Sources: Census of Canada, 2011 and 2016, estimated driving cost per CAA assumptions 2011 and 2016. Driving distance and driving time from Google Maps.**

| Geography | Distance to Toronto (km) | Driving Time (mins.) | Avg 2011 After-Tax House-hold Income ($) | Avg 2016 After-Tax House-hold Income ($) | Avg 2011 Shelter Cost ($) | Avg 2016 Shelter Cost ($) | Total 2011 Driving Costs ($/mo.) | Total 2016 Driving Costs ($/mo.) | Total 2011 Cost ($/mo.) | Total 2016 Cost ($/mo.) | 2011 Shelter +Drive (%) | 2016 Shelter+ Drive (%) | Lone Drivers (2016) |
|---|---|---|---|---|---|---|---|---|---|---|---|---|---|
| Toronto - CMA | 10 | 38 | 78165 | 87993 | 1359 | 1755 | 581.88 | 649.94 | 1940.88 | 2404.94 | 29.8 | 32.8 | 1274835 |
| Oshawa - CMA | 51 | 56 | 75527 | 84163 | 1290 | 1552 | 591.27 | 643.58 | 1881.27 | 2195.58 | 29.89 | 31.3 | 11035 |
| Port Hope - CA | 75 | 67 | 66065 | 74054 | 1022 | 1243 | 761.67 | 804.77 | 1783.67 | 2047.77 | 32.4 | 33.18 | 140 |
| Hamilton - CMA | 77 | 65 | 70277 | 80008 | 1130 | 1422 | 775.87 | 818.2 | 1905.87 | 2240.2 | 32.54 | 33.6 | 17725 |
| St. Catharines - Niagara - CMA | 79 | 87 | 60999 | 67788 | 974 | 1153 | 790.07 | 831.63 | 1764.07 | 1984.63 | 34.7 | 35.13 | 720 |
| Kitchener - Cambridge - Waterloo - CMA | 94 | 89 | 71388 | 79056 | 1140 | 1381 | 896.57 | 932.37 | 2036.57 | 2313.37 | 34.23 | 35.11 | 4550 |
| Centre Wellington - CA | 94 | 80 | 77041 | 82668 | 1185 | 1385 | 896.57 | 932.37 | 2081.57 | 2317.37 | 32.42 | 33.64 | 160 |
| Peterborough - CMA | 95 | 90 | 62509 | 68638 | 1000 | 1202 | 903.67 | 939.09 | 1903.67 | 2141.09 | 36.55 | 37.43 | 350 |
| Orillia - CA | 96 | 91 | 57707 | 61112 | 1009 | 1174 | 910.77 | 945.8 | 1919.77 | 2119.8 | 39.92 | 41.62 | 185 |
| Guelph - CMA | 97 | 82 | 72864 | 82559 | 1181 | 1479 | 917.87 | 952.52 | 2098.87 | 2431.52 | 34.57 | 35.34 | 3400 |
| Woodstock - CA | 97 | 91 | 58874 | 68803 | 1005 | 1193 | 917.87 | 952.52 | 1922.87 | 2145.52 | 39.19 | 37.42 | 115 |
| Ingersoll - CA | 106 | 97 | 63699 | 72019 | 998 | 1232 | 981.77 | 1012.96 | 1979.77 | 2244.96 | 37.3 | 37.41 | 35 |
| Kawartha Lakes - CA | 106 | 100 | 62355 | 69163 | 1009 | 1150 | 981.77 | 1012.96 | 1990.77 | 2162.96 | 38.31 | 37.53 | 260 |
| Cobourg - CA | 108 | 82 | 61560 | 68651 | 1006 | 1189 | 995.97 | 1026.4 | 2001.97 | 2215.4 | 39.02 | 38.72 | 120 |
| Midland - CA | 108 | 99 | 57563 | 63290 | 970 | 1115 | 995.97 | 1026.4 | 1965.97 | 2141.4 | 40.98 | 40.6 | 105 |
| Norfolk - CA | 110 | 118 | 60840 | 67807 | 893 | 1061 | 1010.17 | 1039.83 | 1903.17 | 2100.83 | 37.54 | 37.18 | 95 |



| Stratford - CA | 111 | 116 | 61342 | 67646 | 972 | 1162 | 1017.27 | 1046.54 | 1989.27 | 2208.54 | 38.91 | 39.18 | 70 |
| Collingwood - CA | 112 | 113 | 62016 | 68585 | 1076 | 1255 | 1024.37 | 1053.26 | 2100.37 | 2308.26 | 40.64 | 40.39 | 80 |
| Barrie - CMA | 112.2 | 79 | 70402 | 80832 | 1297 | 1521 | 1025.79 | 1054.6 | 2322.79 | 2575.6 | 39.59 | 38.24 | 3085 |
| Belleville - CA | 123 | 111 | 58520 | 65204 | 939 | 1127 | 1102.47 | 1127.14 | 2041.47 | 2254.14 | 41.86 | 41.48 | 250 |
| Brantford - CMA | 127 | 91 | 62608 | 70718 | 1023 | 1227 | 1130.87 | 1154 | 2153.87 | 2381 | 41.28 | 40.4 | 500 |
| London - CMA | 128 | 121 | 63208 | 69903 | 1007 | 1219 | 1137.97 | 1160.72 | 2144.97 | 2379.72 | 40.72 | 40.85 | 835 |
| Owen Sound - CA | 156 | 142 | 59011 | 64148 | 900 | 1089 | 1336.77 | 1348.76 | 2236.77 | 2437.76 | 45.49 | 45.6 | 50 |

## 4.0 Findings and Results

*4.1 Data Normality and Spatial Autocorrelation.*     Using Minitab 14, the Ryan-Joiner test was used to evaluate normality of the data. Spatial autocorrelation using Global Moran's $I$ was checked using GeoDa 1.18.0 (Anselin et al., 2006). The results are displayed in Table 4 below. Each of the variables do not report significant values of Global Moran's $I$, suggesting that they are the result of random processes, and we are free to use OLS without having to specify a spatial lag or spatial error model (Anselin, 2007, p.199).  The Ryan-Joiner test mostly identified all variables that had normally distributed errors, with the exception of 2011 and 2016 average after-tax household income and average shelter costs which did not meet the criteria of p>0.100. However, this issue disappeared when the variables were used to compute the total cost of shelter and driving costs and the percentage of income devoted to shelter and driving costs.

**Table 4:  Means, Global Moran's I and Ryan-Joiner test statistics**

| Variable | Mean | Global Moran's *I* (p-value) | Ryan-Joiner (p-value) |
| --- | --- | --- | --- |



| 2011 Average After-tax Household Income | $64,980/year | -0.5204 (0.603) | 0.942 (0.022) |
|---|---|---|---|
| 2016 Average After-tax Household Income | $72,383/year | -0.8280 (0.408) | 0.958 (0.068) |
| 2011 Average Shelter Cost | $1060/month | 0.263 (0.793) | 0.939 (0.014) |
| 2016 Average Shelter Cost | $1273/month | 0.126 (0.900) | 0.934 (<0.010) |
| 2011 Total Driving Cost | $11,314/year | 0.009 (0.993) | 0.970 (>0.100) |
| 2016 Total Driving Cost | $11,722/year | 0.009 (0.993) | 0.971 (>0.100) |
| 2011 Total Cost | $2,003/month | 0.985 (0.374) | 0.984 (>0.100) |
| 2016 Total Cost | $2,250/month | 0.616 (0.538) | 0.990 (>0.100) |
| 2011 Shelter + Driving % | 37.10% | 0.183 (0.855) | 0.981 (>0.100) |
| 2016 Shelter + Driving % | 37.57% | 0.068 (0.946) | 0.986 (>0.100) |

Source: Minitab 14 calculations

*4.1 Regression Models.*     Using 2011 and 2016 shelter cost data for census metropolitan areas and census agglomerations within approximately a 156 kilometre radius of Toronto (n=23) and applying OLS regression against the Google driving distance in kilometres of each community to Toronto, shelter cost functions (i.e., ***R(d)***) were estimated for each year. In Table 6, as expected, a negative slope is observed as shelter costs decrease increasing distance from Toronto for both Model 3 and Model 4. In relative terms, there was an increase in the constant for shelter costs from 2011 to 2016, and an increase in the slope parameter, which suggests that the shelter cost curve has not only shifted upward



but also became more steep with increasing distance from Toronto. While both models are statistically significant at both the 1% and 5% levels, the explanatory power of the OLS regressions are not as high as anticipated, although curiously stronger for the 2016 data ($r^2$=0.507) than the 2011 data ($r^2$=0.446). Quadratic regressions ($r^2$=0.457, 0.547 for 2011 and 2016, respectively) and cubic regressions ($r^2$=0.461, 0.55 for 2011 and 2016, respectively) showed marginal improvement in fit, with cubic regression yielding the best results.

The Total Cost (i.e., shelter cost plus driving cost) curves for 2011 and 2016 were also regressed against distance from Toronto **(Table 6, Model 1 and Model 2)**. Similar to the outcome with shelter cost data, cubic regression yielded a better fit versus OLS and quadratic regressions. While most of the Total Cost curve regression models are mostly significant at the 1% and 5% levels, the 2016 OLS results were not significant at either the 1% or 5% levels. An increase in the constant term for Total Cost from 2011 to 2016 was observed, suggesting that the Total Cost curve has shifted upward in 2016. Furthermore, an increase in the slope parameters from 2011 to 2016 were also observed, suggesting that the Total Cost curve in 2016 also became more steep.

As an example of the effects on a household earning $60,000 per year after-taxes, the results of these cost shifts are displayed on Figure 4, given different percentages of income constraints on the household. Income is assumed to be the same in both 2011 and 2016. Figure 4 illustrates the shelter cost, driving cost, and total cost curves for 2011 and 2016, and the curves for 2016 are shifted above those for 2011. Intersecting most of these curves are the income guideline levels which are plotted horizontally. It is clear that the total cost curve for 2016 is well above the 42% income guideline for mortgage qualification, implying that our example household will need to spend a larger percentage of income to afford these new costs at any distance from Toronto, or simply find more income. The consequence of this situation is that the household risks disqualification if it spends a greater



percentage of income. However, in 2011, that same household could live as far out as 120 kilometres at the 42% income guideline. Where the 30% income guidelines intersect the shelter cost curves on Figure 4, the zone of feasible housing has been pushed out due to the shift in shelter costs from 2011 (when the allocated household income was greater than shelter costs) to almost 50 kilometres from Toronto in 2016. This means that shelter costs at distances less than 50 kilometres from Toronto are infeasible for the household, given that shelter costs in this region are higher than the allocated household income. In general, the model can be used to evaluate the effects of cost shifts over time for any income level, and predict the change in commuting distance with a given income allocation.

The distribution of household income as a function of distance may give some insight into the role of income and commuting. Table 5 illustrates regression results of this relationship. Although significant at 0.01, after-tax household income in 2011 ($r^2 = 0.413$) and 2016 ($r^2 = 0.42$) does not have a strong correlation with distance from Toronto. The results suggest that income declines with increasing distance from Toronto. While the income data in this study did not exhibit a convex pattern as a function of distance, similarly noted by Hu et. al (2020), minimums of after-tax household income were observed at 108 kilometres from Toronto in 2011 and 96 kilometres in 2016. In Table 3, it is interesting to note that income levels dramatically increased from 2011 to 2016 in these (and other) communities outside of Toronto (some of which are rural and not industrialized), suggesting that higher income workers may be moving to these areas and commuting.

**Table 5: Model Estimates: After-tax Household Income as a function of distance**

**Source:** regression model estimates

|  | 2011 After-tax Household Income | 2016 After-tax Household Income |
|---|---|---|



| Independent Variable | Distance (km) | Distance (km) |
|---|---|---|
| **Linear Regression** | $r^2=0.413$<br>$p < 0.001$<br>$s= 5028.48$ | $r^2=0.420$<br>$p < 0.001$<br>$s= 5843.77$ |
| **Constant** | 79188<br>($p < 0.001$) | 89129<br>($p < 0.001$) |
| **Slope Parameter** | -143.82d<br>($p=0.001$) | -169.51d<br>($p=0.001$) |

**FIGURE 4: Shelter, driving and budget guideline curves, 2011 and 2016**
**Source: CMHC Mortgage Stress rule percentages, model calculations from dataset**
(color required, 1.5 column)



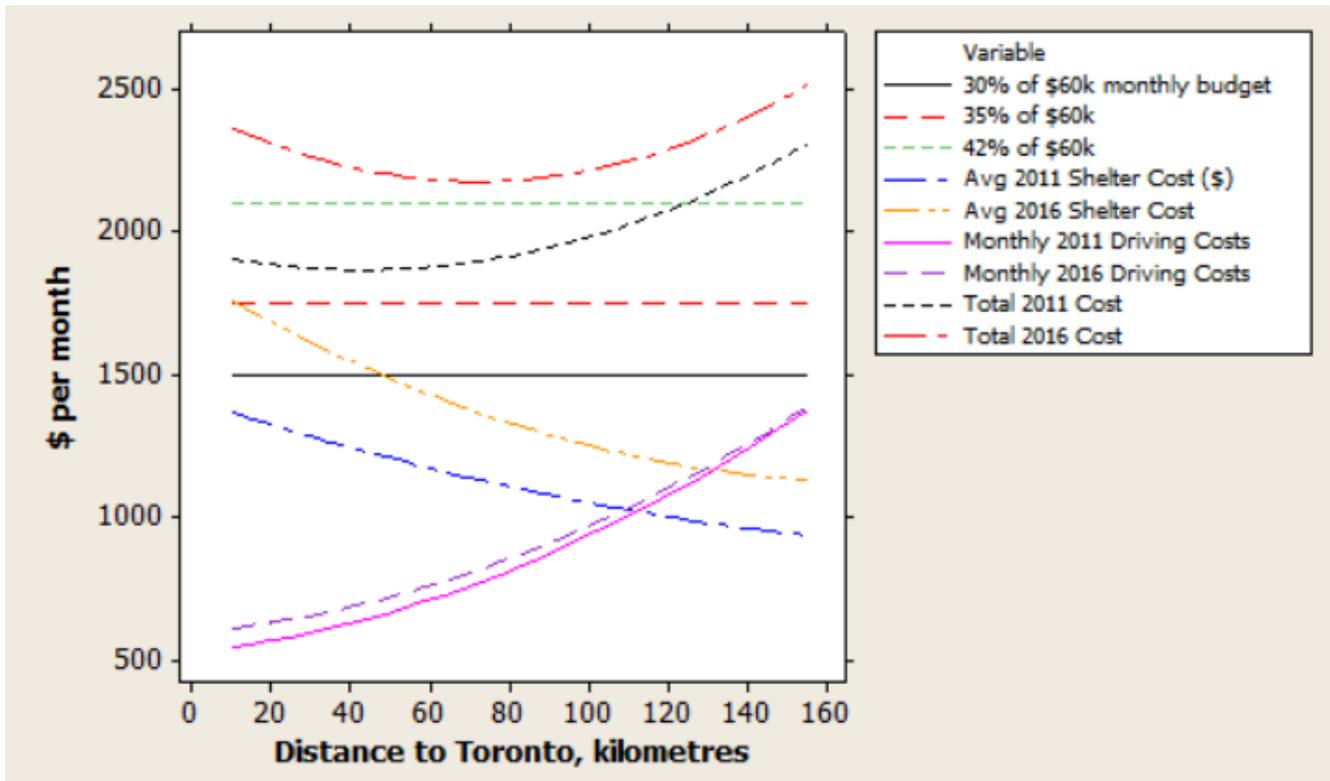

The percentage of household income devoted to both shelter cost and driving cost for 2011 and 2016 were regressed against distance from Toronto. Although all regressions were significant at the 1% and 5% levels, cubic regression yielded the best fit for 2011 and 2016 percentages ($r^2$=0.851, 0.808 for 2011 and 2016, respectively) over OLS and quadratic regression. An increase in the constant term is observed from 2011 to 2016, with 31.33% of household income being required in Toronto in 2011 versus 34.22% in 2016, which approaches the CMHC 35% of household income guideline. These curves are upward sloping with increasing distance from Toronto, and appear to become more steep with distance. This result confirms the observation by Coulombel (2018) that household cost burden increases as distance increases away from the city.

Figure 5 illustrates the percentage of income spent on shelter and transportation versus distance from



Toronto for 2011 and 2016. As with Figure 4, budget guidelines are drawn horizontally to intersect the curves. Both curves clearly begin to converge around a distance of 100 kilometres from Toronto and are essentially the same where the 42% budget guideline intersects these curves (about 130 kilometres). At distances beyond the intersection of the 42% budget guideline and the curve is the region of the model that represents unfeasible housing due to high transportation costs. This also corresponds to what is observed on Figure 4 at about the same distance for the 2011 Total Cost curve (i.e., Total Cost exceeds allocated income at this distance). The traditional 30% of income guideline is illustrated for comparison to the other budget guidelines. The zone between the 35% and 42% budgeted income is the region of feasible housing to qualify for a mortgage, a distance band approximately between 80 and 150 kilometres from Toronto based on Figure 5. The 45% guideline in Figure 5 represents the percentage of income that is regarded as the maximum that could be spent by the household on shelter and transport before being considered unaffordable (Guerra & Kirschen (2016)), and corresponds to a distance of about 150 kilometres from Toronto. Since Table 3 confirms that very few people commute by car from beyond 150 kilometres, this implies that commuters are discouraged from doing so due to this high percentage of income being spent on shelter and transportation.



**Figure 5: Budget guidelines versus shelter + driving percent of income curves, 2011 and 2016
Source: CMHC Mortgage Stress rule percentages, model calculations from dataset**
(color required, 1.5 column)

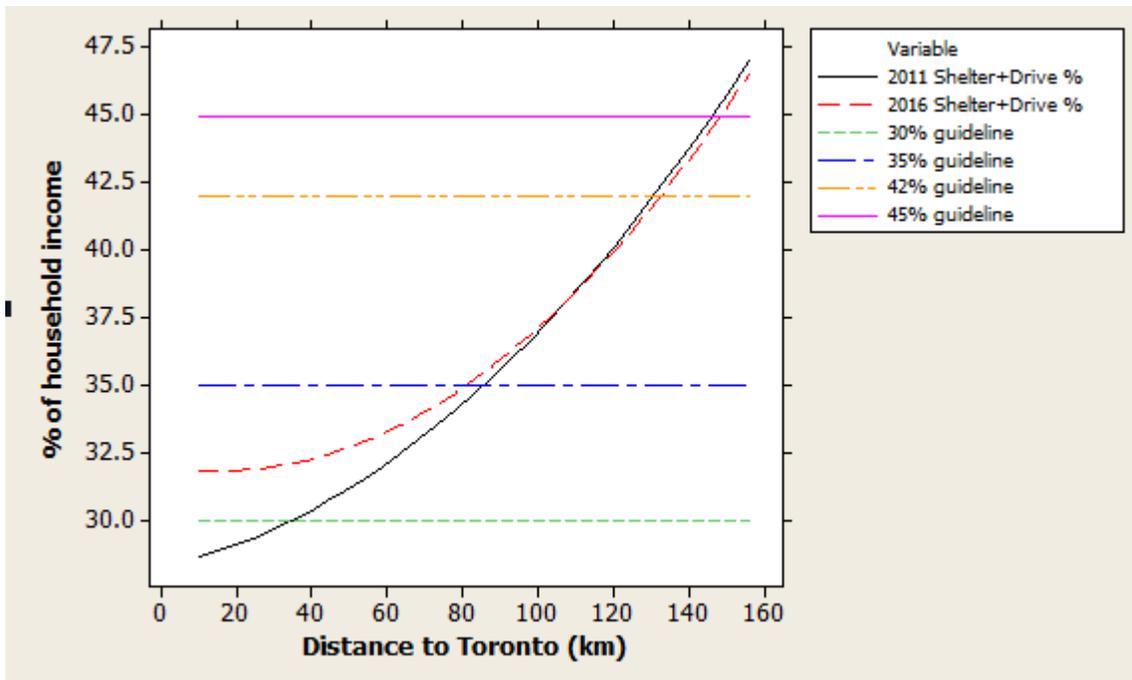

The shelter cost curves for 2011 and 2016 census data were estimated using distance from Toronto (Table 6). The constant term increased from 1353 in 2011 to 1699 in 2016, indicating an overall upward shift in the shelter cost curve for the region during this period due to higher shelter costs. The slope coefficient increased from -2.962 in 2011 to -4.314 in 2016, suggesting that since the shelter cost curve became more steep with distance. The R-squared increased from 0.456 in 2011 to 0.507 to 2016, with the constant and distance terms being significant at 1% and 5%. Table 6 also illustrates that cubic regression yielded a better over OLS and quadratic regression, with R-squared values of 0.461 and 0.55 for 2011 and 2016, respectively. In general, distance from Toronto only explains about half of the variation in shelter costs. Furthermore, the relationship between shelter costs and distance may be



curvilinear, as suggested by Figure 2, and improved fit through quadratic and cubic regression in Table 6.

**Table 6: Model Estimates - Total (Monthly) Cost, Monthly Shelter Cost and Shelter plus Driving Cost as a Percentage of Household Income as a function of distance ($d$)**
**Source:** Regression estimates

| Model | Model 1 | Model 2 | Model 3 | Model 4 | Model 5 | Model 6 |
|---|---|---|---|---|---|---|
| Description | 2011 Total Cost | 2016 Total Cost | 2011 Shelter Cost | 2016 Shelter Cost | 2011 Shelter + Driving Cost as % of Household Income | 2016 Shelter + Driving Cost as % of Household Income |
| Cubic Regression | $r^2$=0.535 p=0.002(*) s=98.5 | $r^2$=0.340 p=0.044(**) s=126.07 | $r^2$=0.461 p=0.007(*) s=100.45 | $r^2$=0.55 p=0.001(*) s=125.44 | $r^2$=0.851 p=0.000(*) s=1.7 | $r^2$=0.808 p=0.000(*) s=1.64 |
| Constant | 2050 | 2554 | 1438 | 1871 | 31.33 | 34.22 |
| Slope Parameters | $-11.38d$ $+ 0.1531d^2$ $-0.000476d^3$ | $-15.75d$ $+0.1736d^2$ $-0.000493d^3$ | $-6.653d$ $+0.03967d^2$ $-0.000119d^3$ | $-10.64d$ $+0.0582d^2$ $-0.000136d^3$ | $-0.1881d$ $+0.003688d^2$ $-0.000012d^3$ | $-0.1822d$ $+0.003023d^2$ $-0.000009d^3$ |
| Quadratic Regression | $r^2$=0.492 p=0.011(**) s=100.31 | $r^2$=0.299 p=0.029(**) s=124.63 | $r^2$=0.458 p=0.002(*) s=98.18 | $r^2$=0.547 p=0.000(*) s=122.54 | $r^2$=0.820 p=0,000(*) s=1.82 | $r^2$=0.785 p=0.000(*) s=1.69 |
| Constant | 1932 | 2432 | 1409 | 1867 | 28.31 | 32.01 |
| Slope Parameters | $-2.989d$ $+0.03473d^2$ | $-7.049d$ $+0.04878d^2$ | $-4.56d$ $+0.00967d^2$ | $-8.237d$ $+0.02372d^2$ | $0.02786d$ $+0.000591d^2$ | $-0.02374d$ $+0.00075d^2$ |
| Linear Regression | $r^2$=0.347 p=0.003(*) s=111 | $r^2$=0.042 p=0.348 s=142.10 | $r^2$=0.456 p=0.000(*) s=96.91 | $r^2$=0.507 p=0.000(*) s=124.81 | $r^2$=0.775 p=0.000(*) s=1.99 | $r^2$=0.684 p=0.000(*) s=2.00 |
| Constant | 1731 | 2149 | 1353 | 1699 | 24.89 | 27.66 |
| Slope Parameter | $2.755d$ | $1.019d$ | $-2.962d$ | $-4.314d$ | $0.1257d$ | $0.1004d$ |



(*) significant at 1%
(**) significant at 5%

**Table 7: Model Estimates: Monthly Driving Costs**
**Source:** regression model estimates

|  | **2011 Monthly Driving Costs** | **2016 Monthly Driving Costs** |
|---|---|---|
| **Linear Regression** | $r^2$=0.932 p=0.000(*) s= 44.36 | $r^2$=0.926 p=0.000(*) s= 44.36 |
| **Constant** | 378.06 | 499.96 |
| **Slope Parameter** | 5.7169d | 5.3329d |

Monthly driving costs for 2011 and 2016 were modeled as a function of car driving distance to Toronto. The R-squared values reduced slightly from 0.932 to 0.926 for 2011 and 2016, respectively. The regressions and variables were significant at the 1% and 5% levels. There was a definitive increase in monthly driving cost as illustrated by the change in the constant term from $378.06 in 2011 to $499.96 in 2016. The change in the constant term between 2011 and 2016 can be generally attributed mainly to increases in insurance, licensing and finance costs (see Table 2). The driving cost per kilometre (as given by the slope parameter) went down from $5.72 in 2011 to $5.33 in 2016, which we would generally attribute to improvements in fuel efficiency and lower gas prices in 2016 (see Table 2). However, the change in monthly driving costs in Table 7 is clearly overshadowed by the increase in monthly shelter costs during the same period.



*4.3 Limits of Commuting Distance.*     The number of lone car drivers to Toronto (Table 3) may

provide us with some insight into the exodus beyond the suburbs in search of affordable housing. Using

data from the 2016 Journey to Work (workplace city is Toronto) in the dataset illustrates that the

volume of lone car drivers becomes quite low with distance, although there are large contingents of car

commuters in cities as far away as London. The observation of these large numbers of lone car drivers

at great distances suggests that people are making deliberate choices to live where they do, and perhaps

at great cost. Perhaps their location may be related to their overall level of household income. While

the data collected by Statistics Canada does not provide us with any insight about the income levels of

these drivers in the Journey to Work data, generalizations can be made using census data for household

income in these communities. Table 3 does generally imply that as income increases, the number of car

commuters also increases.

Table 8 illustrates the limits of commuting distances based on income allocation at 42% and 45%.

Since the total cost exceeds the allocation of income for households earning annual incomes of

$30,000 and $40,000, they essentially do not have the option of commuting or living at a great

distance. As costs increased from 2011 to 2016, the maximum limit of commuting distance at the 42%

level (i.e., risking mortgage disqualification) and 45% level (i.e., unaffordability) decreased. For

example, in 2011, a household earning $60,000 per year could afford to live no further than 188

kilometres from Toronto using an allocation of 45% of household income. However, in 2016, this same

household income would be restricted to 92 kilometres due to the upward shift in costs. Thus, as

households search further afield for less expensive housing, the increase in costs over time eventually

will result in the erosion of any locational advantage, as shown by the reduction in the maximum

commuting distances in Table 8.



**Table 8: Estimates of Limits of Commuting Distances (kilometres from Toronto) based on 42% and 45% Allocation of Monthly Income, 2011 and 2016**

**Source: Model calculations**

| Annual Household Income | $30,000 | $40,000 | $50,000 | $60,000 |
|---|---|---|---|---|
| **42% Allocation of Monthly Income** | **$1,050** | **$1,400** | **$1,750** | **$2,100** |
| **45% Allocation of Monthly Income** | **$1,125** | **$1,500** | **$1,875** | **$2,250** |
| **Commuting Limit, km from Toronto: 42% of income (2011)** | * | * | 7 | 134 |
| **Commuting Limit, km from Toronto: 42% of income (2016)** | * | * | * | * |
| **Commuting Limit for Commuting, km from Toronto: 45% of income (2011)** | * | * | 52 | 188 |
| **Commuting Limit for Commuting, km from Toronto: 45% of income (2016)** | * | * | * | 92 |

(*) commuter is constrained by high shelter and transport costs at 45% of income - not affordable

**5. Conclusion.**    Using census shelter cost data as a function of distance explains about half of the variation in average shelter costs for communities that are less than 156 km from Toronto. However, additional examination of the data reveals a model of shelter cost and driving cost as a percentage of after-tax household income using distance as an independent variable that explains 80.8% of the variation in this percentage, and delineate zones of feasible and infeasible housing based on various income guidelines.  We conclude that the 35% GDS ratio, as recommended by the CMHC, is clearly below the existing level of expenditure on shelter and transport, The 35% GDS ratio may be feasible only for households earning higher levels of income than those at lower levels of income that spend a higher percentage on shelter and transport.  Since upward shifts in shelter costs from 2011 to 2016 were more pronounced than the changes in driving costs during the same period, the effect was to make the cost of living in (or close to) Toronto impossible to afford for certain income groups. The result of this



effect is to continue the drive until you qualify phenomenon. At 42% of household income, the changes in costs between 2011 and 2016 resulted in a reduction in the maximum commuting distance that a household should not exceed under mortgage stress rules (i.e., 42% of household income).

## 6.0 Implications for Managerial Practice

The effects of urban housing costs (and changes in these costs over time) in major cities have implications not just for those employed in metropolitan areas but also job seekers who may seek to relocate to cities with overheated housing markets. The result is population growth in areas that may lack public intercity transportation infrastructure to curb the growth in car commuting. The shifts in housing costs that create the growth of housing developments aimed at disaffected big-city home buyers in far-flung communities should signal a future requirement to anticipate and plan to overcome future local transport infrastructure strains (e.g., traffic problems, accidents, demands for commuter rail, etc.). Thus, the model can help provide a framework for those who study changes in housing and transport costs for a metropolitan region and predict the changes in commuting and household expenditure based on income level. Future research could also observe the changes in household income in communities, particularly those that have traditionally experienced economic downturn or deindustrialization, as a result of increased numbers of drive until you qualify car commuters moving to another city. The research should also focus on the phenomena of drive until you qualify and the financial well-being of house buying car commuters in overheated urban housing markets. More importantly, attributing commuting behaviors to income data and household characteristics of those that drive until you qualify will help fill the shortcomings of using census data.

The social impact of spending larger proportions of household income on shelter and transportation than is recommended should also be cause for concern. Those households with modest to lower



incomes are not just at risk of being priced out of a local market but also at risk of not being able to sustain higher than accepted levels of spending on shelter and transportation, especially with increases in these costs over time that were illustrated in this paper. There is no question that there would be stress upon commuters and negative effects upon the overall well-being of communities that have large contingents of commuters, including the quality of life in their new community.  More importantly, many people make the decision to either move closer to a city with high shelter costs to be closer to employment (due to limited local employment opportunities) or commute from their home community (which may be at a lower shelter cost) to retain social ties. Directions for future research using this model can be enabled to explore such move-or-commute decisions, especially for areas that have poor transport connectivity and poor local employment market conditions (e.g., cities with high unemployment or declining industrial bases that rely on neighboring cities for employment).

As a strategic tool to gravitate car drivers to transit in distant areas, the transport cost component part of the model can be used to frame a viable transit-friendly value proposition to car commuters and get them to substitute transit use instead. This acknowledges that households which must commute from greater distances need to solve this cost problem. In general, if car drivers prefer to drive because the monthly cost is less than the monthly fare for transit (or a combination of driving and transit), transit planners should use the model to estimate driving costs to come up with an attractive and competitive fare scheme that is more economical than driving. The result should be increased revenues and ridership for the transit operator, with society enjoying a benefit from less gridlock and pollution due to fewer car commuters, and the household having reduced transport expenses.

**7.0 Contribution to Scholarly Knowledge**

The purpose of this article was to present a novel way to model commuting distance by adapting elements of traditional mainstream microeconomics, given an income, a regional housing market and a



proportion of income allocated to shelter and driving. The model presented is an intuitive and elegant identity that can be graphically represented for easy analysis to generalize the anticipated effects upon commuting distance due to shifts in shelter cost, transport cost and income changes for a region. The article also examines the differences in regression methods to estimate shelter cost curves, as well as the shortcomings of assuming that shelter costs are strictly linear in nature for all given distances in a region. Furthermore, as previously described in section *3.2*, the model allows the flexibility of adding additional modes of transport other than car commuting (or in combination) to potentially examine the effect of green modes of transport or alternative commuting strategies on journey to work distance (e.g., cost of ridesharing or carpooling, driving from a distant residence to a regional commuter bus or train station). While the model does not suggest a sole or unique locational decision for a commuter, it does help suggest the limits of commuting, given an income level and percentage of income allocated to shelter and driving cost. The model also provides a starting point for evaluating alternatives with respect to budget constraints (i.e., percentage of income guidelines) that favour better cost control for the household, and the feasibility of living at a given distance with limited transport connectivity. The model offers a new perspective and intuitive approach to examine the problems of urban growth and the increase of car commuting using publicly-available secondary data, with little derivation required.

## 8.0 Limitations

The limitations to this study generally relate to availability of data, and issues that may occur when the model is used in certain situations. Since data for census agglomerations (CAs) and census metropolitan areas (CMAs) in southern Ontario were used to model the cost curves for the Toronto region, there were shortcomings in using this approach. First, since car commuting costs are not reported in Canadian census data at any level of geography (i.e., CA, CMA), overcoming this limitation required the estimation of commuting costs using costing data from secondary sources (i.e., CAA, etc.). Second, since the commuting data from the Census of Canada does not report the mode of travel by



income level, which would be useful for any researcher studying commuting by wage groups, more detailed data would have to be collected through surveys in order to overcome the limitation. Third, within an intra-zonal context, great care must be given to the attention to the potential for the small scale of distance. While the paper presented here illustrates longer distance commuting, smaller distance commutes on an intra-zonal basis may, instead, prefer to use minutes as a distance proxy.